\documentclass[onecolumn]{revtex4}

\usepackage{amssymb}
\usepackage{amsmath}
\usepackage[utf8]{inputenc}
\usepackage{graphicx}
\usepackage{dsfont}
\usepackage{comment}
\usepackage{xcolor}

\def\bra#1{\mathinner{\langle{#1}|}}
\def\ket#1{\mathinner{|{#1}\rangle}}
\def\braket#1{\mathinner{\langle{#1}\rangle}}

\begin{document}
	
	\title{Deterministic multi-mode nonlinear coupling for quantum circuits}
	
	\author{Seckin Sefi, Petr Marek, Radim Filip}
    \affiliation{Department of Optics, Palacky University, 17. listopadu 1192/12, 77146 Olomouc, Czech Republic}
	\email{secgin.sefi@upol.cz}

	\begin{abstract}
We present a general technique for deterministically implementing a multi-mode nonlinear coupling between several propagating microwave or optical modes in quantum circuits. The measurement induced technique combines specifically prepared resource states together with feasible feed-forward operations. We explore several ways of generating the suitable resource states and discuss their difference on an illustrative example of cubic coupling between two modes. We also show  that the required entangled states with requisite nonlinear properties can be already generated in the present day experiments.
	\end{abstract}
	
	\date{\today}
	\maketitle

\section{Introduction}
The fundamental practical applicability of quantum systems largely depends on the dimension of the Hilbert spaces used for their description. Quantum systems with large dimension are prohibitively difficult to classically simulate but, in turn, their controlled dynamics can provide us with answers we cannot reach classically, be it for the lofty quantum computation \cite{Feynman1982,Lloyd1996}, or the more approachable quantum simulation \cite{Buluta2009, Trabesinger2012,Zhang2017} or metrology \cite{Giovannetti2011,Safranek2016,Oh2017}. Large dimension can be obtained by parallelization - combining several quantum systems and treating them as one. The most straightforward example are the recently applied multi-qubit architectures with photons \cite{Zhong2018}, trapped ions \cite{Lanyon2011}, or superconducting systems \cite{Barends2015}.
A complementary approach uses infinite dimensional Hilbert space of quantum modes propagating in circuits to reduce necessary parallelism. However, to optimally utilize this potential, even the already infinite dimensional harmonic oscillators can benefit from multi-oscillator architecture \cite{CVclusters,Cai2017}.

The processing power of such circuit networks is largely given by the amount of control we can exert over the systems. The control comes in two broad varieties - local manipulation of individual modes and interaction coupling of different modes. Their availability is given by the physical nature of the harmonic oscillators. For systems in the form of electromagnetic field modes at microwave or optical frequencies, the local operations start as simple time delay elements that introduce phase shifts and progress up to to linear quantum amplifiers \cite{Weedbrook2012}. Similarly, the multi-mode interactions start with a basic linear coupler, for light represented by a partially transparent mirror, and again progress up to linear multi-mode interactions capable of generating Gaussian quantum entanglement \cite{Weedbrook2012}. These elementary linear operations can be realized taking advantage of existing materials. However, to gain the full benefit of technologies based on quantum systems we need to be able to implement operations that are far from linear. 

In principle, an arbitrary, even highly nonlinear, operation on modes of propagating radiation can be decomposed into into a sequence of elementary unitary gates \cite{Lloyd1999,Sefi2011}. Majority of these gates implement Gaussian operations, both single-mode and two-mode \cite{Weedbrook2012}. The Gaussian operations can be considered well accessible in contemporary experiments; their largest disadvantage is that alone they are not sufficient \cite{Mari2012}. To generate an arbitrary operation we also need access to non-Gaussian gates that employ nonlinearity of higher order in order to deterministically transform pure Gaussian states into pure non-Gaussian ones. One of the simplest gates of this kind is the cubic gate which implements an unitary operation with Hamiltonian proportional to the third power of a quadrature operator. An individual cubic gate can be implemented in the measurement induced fashion \cite{GPK} in which the required nonlinearity is imprinted on the target system through Gaussian interaction with a suitably prepared ancilla and subsequent feed-forward. The biggest obstacle is the availability of the requisite ancillary states. They can be prepared approximately \cite{Marek2011,Yukawa2013,Yukawa2013b, miyata2016cubic} but these approximations are imperfect add unwanted noise during implementing the operation. This extra noise makes repeated application of the gates impractical. For single mode operations the number of gates can be reduced by directly implementing non-Gaussian operations of higher order \cite{marek2018general}. Our aim is to provide similar methodology for non-Gaussian interactions between the propagating modes and thus cover all possible complex quantum circuits with the infinite Hilbert spaces.

In this paper we present a general description of a class of non-Gaussian couplings between several propagating field modes together with a measurement induced method for experimental implementation in quantum circuits. The method is a final generalization of the original GKP proposal \cite{GPK}. The architecture of the $N$ mode coupling is based on $N$ sequential Gaussian quantum nondemolition interactions with $N$ auxiliary modes prepared in a specific nonlinear state, homodyne detection of the ancillary modes, and nonlinear feedforward control applied to the modes that remain.
We consider a third-order non-Gaussian coupling as an example and discuss several different approaches towards the entangled resource state generation. Ultimately we show that genuinely non-Gaussian coupling coupling between several harmonic oscillators is feasible, which opens up the avenue of both quantum technologies and fundamental research of highly nonlinear quantum mechanics and thermodynamics.



\section{Multi-mode non-Gaussian coupling}
A general unitary coupling between $N$ modes can be characterized by interaction Hamiltonian expressed as a functional $V_{xp}(\hat{x}_1,\hat{p}_1,\cdots,\hat{x}_N,\hat{p}_N)$, where $\hat{x}_j$ and $\hat{p}_j$ are quadrature operators of the individual modes with commutation relations $[\hat{x}_j,\hat{p}_k] = i\delta_{jk}$.The required Hermitian nature of the Hamiltonian demands the underlying real function $V_{xp}(.)$ to be symmetrical with respect to exchanging operators $\hat{x}_j$ and $\hat{p}_j$ for all $j$. Implementation of such general operation is still an open problem. However, the mono-quadrature subclass consisting of operation with interaction Hamiltonians that are real functions of only single type of quadrature operators, $V_x(\hat{x}_1,\cdots,\hat{x}_N)$ can be implemented by generalizing the original proposal for cubic gates \cite{GPK}. Hamiltonians of such operations are implicitly Hermitian and no further constrains on the form of the function are required. When used as part of larger circuits over the elementary single mode cubit gates, these more complex operations can significantly decrease the overall number of gates required and thus simplify the realization of quantum circuits \cite{Sefi2011,marek2018general}.

\begin{figure}
    \centering
\includegraphics[width=0.65\textwidth]{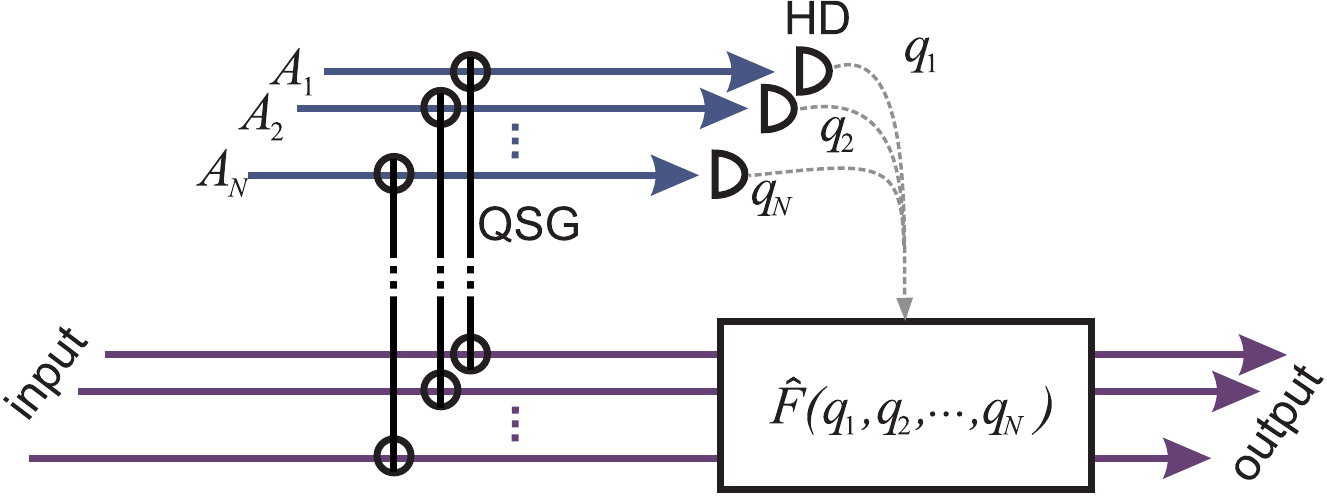}
    \caption{Schematic depiction of the measurement induced implementation of a mono-quadrature multi-mode coupling. The input modes interact with the auxiliary modes $A_1,\cdots, A_N$ through a sequence of quadrature sum gates (QSG). The auxiliary modes are then measured by homodyne detectors (HD), yielding values $q_1,\cdots,q_N$, which are then used for the corrective feed-forward $\hat{F}(q_1,\cdots,q_N)$, which is a shorthand notation for (\ref{Ucorrective}), in order to produce the output.}
    \label{fig:setup}
\end{figure}

Any operation from this class transforms an unknown quantum state $|\psi\rangle$ into $e^{-iV_x(\hat{x}_1,\cdots,\hat{x}_N)}$. It can be implemented by the measurement induced circuit schematically depicted in Fig.~\ref{fig:setup}. In the circuit, the $N$ propagating field modes in an unknown state
\begin{equation}\label{}
   |\psi\rangle = \int_{\mathbb{R}^N} \psi(x_1,\cdots,x_N)|x_1,\cdots,x_N\rangle dx_1 \cdots dx_N
\end{equation}
interact with $N$ auxiliary modes prepared in a specific resource state
\begin{equation}\label{ancilla}
    |\mathcal{A}\rangle = \int_{\mathbb{R}^N} e^{-i V(y_1,\cdots,y_N)} |y_1,\cdots,y_N\rangle dy_1 \cdots dy_N.
\end{equation}
These interactions are local, coupling only two modes at a time by the non-demolition quadrature sum gates (QSG) \cite{QND_gate} characterized by interaction Hamiltonians $\hat{H}_j = \hat{x}_{T,j}\hat{p}_{A,j}$, where the subscripts $T$ and $A$ denote the target and the ancilla, respectively. The effective strength of the coupling does not affect the principal performance of the gate and we will therefore consider it to be equal to one.  After the interactions which transform the joint state of all modes into:
\begin{equation}\label{}
    \int_{\mathbb{R}^{2N}} \psi(x_1,\cdots,x_N) e^{-i V_x(y_1,\cdots,y_N)} |x_1,\cdots,x_N\rangle |y_1-x_1,\cdots,y_N-x_N\rangle  dx_1 \cdots dx_N dy_1 \cdots dy_N
\end{equation}
the $\hat{x}_j$ quadratures of the auxiliary modes are measured, yielding a set of classical values $q_j$. For each set of such values, the joint state of the target modes can be expressed as:
\begin{equation}\label{state_out}
    e^{-i F(\hat{x}_1,\cdots,\hat{x}_N;q_1,\cdots,q_N)} e^{-iV_x(\hat{x}_1,\cdots,\hat{x}_N)}|\psi\rangle.
\end{equation}
Up to the unitary operation $e^{-i F(\hat{x}_1,\cdots,\hat{x}_N;q_1,\cdots,q_N)}$ depending on the values obtained by the measurement, this is the ideal transformed state we would expect to obtain after application of the multi-mode coupling. The measurement dependent operation is represented by effective Hamiltonian
\begin{align}\label{feed-forward}
    F(\hat{x}_1,\cdots,\hat{x}_N;q_1,\cdots,q_N) = V_x(\hat{x}_1+q_1,\cdots,\hat{x}_N+q_N) - V_x(\hat{x}_1,\cdots,\hat{x}_N) \nonumber \\
    =\sum_{\substack{k_1,\cdots,k_N = 0 \\ \sum k_j > 0}  }^{\infty}\frac{q_1^{k_1}\cdots q_N^{k_N}}{k_1!\cdots k_N!}\frac{\partial^{k_1}}{\partial \hat{x}_1^{k_1}}\cdots\frac{\partial^{k_N}}{\partial \hat{x}_N^{k_N}}V_x(\hat{x}_1,\cdots \hat{x}_N).
\end{align}
This undesired term can be compensated in several ways. We can accept only those realizations in which all the measurements yielded values sufficiently close to zero. Such post-selection would result in perfect imprint of the multi-mode nonlinearity of the ancilla on the joint state of the target oscillators, albeit at the cost of exponentially diminishing success rate. Since the low success rate can be an issue in some quantum processing applications, it is worthwhile to find ways to compensate the term deterministically. When the multi-mode interaction Hamiltonian can be expressed as a finite polynomial of the individual quadrature operators, the term can be removed by applying unitary operator
\begin{equation}\label{Ucorrective}
    \hat{U}_{cor} = e^{i F(\hat{x}_1,\cdots,\hat{x}_N;q_1,\cdots,q_N)}.
\end{equation}
The required feed-forward is of the same nature as the operation we are attempting to implement, only with lower order. If the initial attempted operation has an effective Hamiltonian that can be expressed as finite polynomial of the quadrature operators,
\begin{equation}\label{finite_polynomial}
    V_x(\hat{x}_1,\cdots,\hat{x}_{N}) = \sum_{k_1,\cdots, k_N = 0}^{M_1,\cdots,M_N} c_{k_1,\cdots,k_N} \hat{x}_1^{k_1}\cdots \hat{x}_N^{k_N},
\end{equation}
with total maximal power equal to $M_1 +\cdots + M_N = M$, then the corrective operation has an effective Hamiltonian that is different, but of the same form (\ref{finite_polynomial}) only with total maximal power $M_1 +\cdots + M_N = M-1$. Due to this similarity, the same kind of circuit can be employed for implementing the corrective operations, possibly merging the required feed-forwards for the future steps, similarly as in the single mode scenario \cite{marek2018general}.

It should be noted that for the quadrature sum gates can be in experiments replaced by a passive linear couplings - beam splitters. Beam splitters transform the $\hat{x}$ representation kets as $|x_j,y_j\rangle \rightarrow |(x_j+y_j)/\sqrt{2},(y_j - x_j)/\sqrt{2}\rangle$. After the measurement yields value $q_j = (y_j - x_j)/\sqrt{2}$, the ket can be found in the form $|x_j\sqrt{2} + q_j\rangle$ while the corresponding wave functions remain identical. A form equivalent to (\ref{state_out}) can now be obtained by squeezing each mode by factor $\sqrt{2}$ and displacing its $\hat{x}_j$ quadrature by $-q_j/\sqrt{2}$.

The crucial component of the gate is the joint state of the auxillary modes (\ref{ancilla}). The ideal resource state can be approached only asymptotically, which is a feature these states share with the quadrature squeezed states. The exact role of this resource state becomes apparent when we describe the operation in the Heisenberg picture. For the interaction Hamiltonian $V(\hat{x}_1,\cdots,\hat{x}_N)$ the quadrature operators of the target modes are supposed to evolve as:
\begin{equation}\label{}
    \hat{x}_j \rightarrow \hat{x}_j, \quad \hat{p}_j \rightarrow \hat{p}_j - \frac{\partial}{\partial \hat{x}_j}V(\hat{x}_1,\cdots,\hat{x}_N).
\end{equation}
In the measurement induced scheme, the non-demolition nature of the QSG ensures the desired preservation of the $\hat{x}_T$ operators, while the $\hat{p}_T$ operators evolve into $\hat{p}'_{T,j} = \hat{p}_{T,j} + \hat{p}_{A,j}$. The corrective feed-forward now relies on transforming the $\hat{p}_T$ quadratures by applying the unitary operator (\ref{Ucorrective}):
\begin{align}\label{heisenberg_out}
    \hat{p}''_{T,j} &= \hat{p}'_{T,j} + \frac{\partial}{\partial \hat{x}_{T,j}} F(\hat{x}_{T,1},\cdots,\hat{x}_{T,N};q_1,\cdots,q_N) \nonumber \\
    &=  \hat{p}_{T,j} - \frac{\partial}{\partial \hat{x}_{T,j}}V(\hat{x}_{T,1},\cdots,\hat{x}_{T,N}) + \hat{p}_{A,j} + \frac{\partial}{\partial \hat{x}_{A,j}}V(\hat{x}_{A,1},\cdots,\hat{x}_{A,N}).
\end{align}
Here $q_j = \hat{x}_{A,j} - \hat{x}_{T,j}$ are the measured values and we can see that the quadratures are transformed in the manner we desire but the imperfect ancilla contributes by $N$ noise terms, one for each of the modes. Fortunately these noise terms commute and their variances can be, in principle, simultaneously reduced to zero, which is exactly the case of the ideal state (\ref{ancilla}).
Complete removal of the noise terms requires states with infinite energy and thus can be accomplished only asymptotically. This is the long term goal. However. for the short term experimental tests we need to find states which, under feasible and realistic constraints, minimize the variances of these operators and thus produce a high order nonlinear squeezed and entangled states suitable for implementing the operations with a better-then-gaussian performance.

\section{Resource states for the two-mode cubic gate}
Let us now discuss the preparation of the nonlinear resource states. Such states are required to reduce the added fluctuations and enable implementations of the multi-mode gates that are observably superior to the classical approaches. The defining property of such states is that the joint variance of the nonlinear quadratures (\ref{heisenberg_out}
\begin{equation}\label{}
    \sum_{j = 1}^N \langle \{ \Delta [\hat{p}_{A,j} +  \frac{\partial}{\partial \hat{x}_{A,j}}V(\hat{x}_{A,1},\cdots,\hat{x}_{A,N})]\}^2\rangle
\end{equation}
is minimal. Here subscript $A$ marks the operators as belonging to the auxiliary systems but, since we are going to be dealing solely with the resource states from now on, we will be omitting it in the following text. Let us discuss the techniques for minimization on a concrete example of ancillary states for nonlinear coupling with Hamiltonian $\hat{H} = \kappa \hat{x}_1\hat{x}_2^2$. This is operation of the third order, which is the lowest order the operation can have and still be non-Gaussian. The operation has the potential of being experimentally tested in the foreseeable future and, from a practical point of view, it can be used for implementing nonlinear non-demolition gates such as a photon number resolving measurement \cite{sefi2013squeezing}.
Since the operation is of the third order, the corrective feed-forward operations (\ref{feed-forward}) will be of the second order and therefore Gaussian. The only source of non-Gaussian nonlinear behavior is therefore the resource state which should exhibit squeezing in nonlinear quadratures
\begin{equation}\label{eq:quadratures}
   \hat{p}'_{1} =  \hat{p}_{1} + \kappa \hat{x}_{2}^2,\quad \hat{p}'_{2} = \hat{p}_{2} + 2\kappa \hat{x}_{1}\hat{x}_{2}.
\end{equation}
The desired ancillary states are those that, under chosen conditions minimize the sum of the variances of the the nonlocal quadrature operators
\begin{align}\label{minvariance}
   \langle (\Delta\hat{p}'_1)^2\rangle + \langle (\Delta\hat{p}'_2)^2\rangle   =  \bra{\psi}\left(\left(\hat{p}_1+\kappa \hat{x}_2^2\right)^2-d_1^2+\left( \hat{p}_2+2\kappa \hat{x}_1\hat{x}_2\right)^2-d_2^2\right)\ket{\psi},
\end{align}
where $d_1=\bra{\psi}(\hat{p}_1+\kappa \hat{x}_2^2)\ket{\psi}$ and $d_2=\bra{\psi}(\hat{p}_2+2\kappa \hat{x}_1\hat{x}_2)\ket{\psi}$. This moment is the balanced sum of the variances of the nonlinear combination of quadratures (\ref{eq:quadratures}) and is therefore a nonlinear functional of the states. For the ideal state (\ref{ancilla}) the term is equal to zero but for any physical state it serve as a source of added noise in the circuit. Both the quadratures are nonlinear combination of quadratures of different modes. As  a consequence the resource quantum state should exhibit non-Gaussian quantum entanglement manifesting as nonlocal nonlinear squeezing.

The minimization can be separated into two steps. The ancillary states can be expressed as a non-Gaussian core state prepared as a finite superposition of Fock states, which is subsequently adjusted by an optimized two-mode Gaussian operation,
\begin{equation}\label{approximate_ancilla}
    |\psi\rangle = \hat{U}_G \sum_{m=0,n=0}^{M,N} c_{mn} |m,n\rangle.
\end{equation}
This is a feasible two-mode extension of the core state principle \cite{Menzies2009, Huang2015}.
With the help of the Bloch-Messiah decomposition \cite{Braunstein2005} the Gaussian operation on two oscillators can be decomposed into two passive linear coupling operations with two squeezing operations sandwiched in between, together with two displacement operations at the output. This amounts to 12 free parameters which need to be considered in the optimization. Fortunately, some of these parameters can be removed due to the symmetry of the problem. Displacement of $\hat{p}$ quadratures can be neglected because it does not affect the variances. Similarly, the optimal displacement of the $\hat{x}$ quadratures is zero because the operators (\ref{eq:quadratures}) are symmetric with respect to $\hat{x}$. The linear coupling operations can be decomposed into rotation operations and beam splitter operations  \cite{Reck1994}. The nonlinear quadratures (\ref{eq:quadratures}) behave quite differently with regards to their $\hat{x}$ and $\hat{p}$ operators. As a consequence, the rotation operations which mix the different single oscillator quadratures do not contribute towards reduction of the variance. The final Gaussian operation can be therefore composed as
\begin{equation}\label{gaussian_processing}
    \hat{U}_G = \hat{U}_{BS}(\theta_2) \hat{S}_1(\lambda_1)\hat{S}_2(\lambda_2) \hat{U}_{BS}(\theta_1),
\end{equation}
where
\begin{equation}
\hat{U}_{BS}^\dag(\theta)\begin{pmatrix}
    \hat{x}_1 \\ \hat{x}_2 \\ \hat{p}_1 \\ \hat{p}_2
     \end{pmatrix}\hat{U}_{BS}(\theta)= \begin{pmatrix}
    \cos{\theta} &  \sin{\theta} &0 &0 \\ -\sin{\theta} & \cos{\theta} &0 &0 \\ 0&0 & \cos{\theta} & \sin{\theta} \\ 0&0 &-\sin{\theta} &\cos{\theta}
     \end{pmatrix}\begin{pmatrix}
    \hat{x}_1 \\ \hat{x}_2 \\ \hat{p}_1 \\ \hat{p}_2
     \end{pmatrix},
\end{equation}
and
\begin{equation}
    \hat{S}_i^\dag(\lambda_i)\begin{pmatrix} \hat{x}_i\\\hat{p}_i
    \end{pmatrix}\hat{S}_i(\lambda_i)=\begin{pmatrix} \hat{x}_i/\lambda_i\\  \hat{p}_i \lambda_i
    \end{pmatrix}
\end{equation}
represent the composing linear coupling and squeezing operations, respectively. The optimal Gaussian processing therefore has only four relevant parameters, beam splitter parameters $\theta \in (-\pi,\pi]$, and squeezing parameters $\lambda_1, \lambda_2 \in (0,\infty)$. Note that even though we neglect arbitrary phase shifts, by admitting the full $2\pi$ range of the beam splitter parameters we allow for phase flips.
When limited to Gaussian resources the input will be the vacuum state $\ket{\psi}_{12}=\ket{0}_{1}\ket{0}_{2}$ and we will denote the minimal variance (\ref{minvariance}) that can be achieved with such resources by $V_G$. This variance can be found numerically. By introducing non-Gaussian input state we aim to reduce this variance. We will quantify the reduction by using the relative variance $R_V = \frac{V_{NG}}{V_G}$, which is defined as the ratio between the variances of the non-Gaussian and the optimal Gaussian ancillary states. 

Let us start by checking how easily can the nonlinearity required for the gate be created or, in other words, what is the easiest setup that can produce suitable approximation of nonlinear state (\ref{ancilla}). For this we can consider the simplest possible non-Gaussian core states, which are the factorized states of the two oscillators, and reduce the Gaussian processing only to single passive linear coupling. This reduces the difficulty of the numerical optimization as well as the potential experimental implementation. Thus obtained resource states of the form
\begin{equation}\label{simplified}
    |\psi_{M,N}\rangle = \hat{U}_{BS}(t) \sum_{m=0}^M c_m |m\rangle \otimes \sum_{n= 0}^N c'_n|n\rangle
\end{equation}
can be prepared in currently available experimental circuits \cite{Yukawa2012,Pfaff2017} and therefore offer the possibility of experimental verification. The results of the numerical optimization, which is in detail described in Appendix A, are summarized in Fig.~\ref{fig:singlemode_no_squeezing_comp}. The relative variances of the auxiliary states are plotted in relation to the value of the nonlinear parameter $\kappa$. We can see that nonlocal nonlinear squeezing which surpasses the optimal Gaussian states with arbitrary squeezing can be constructed even with the limited resources we consider. Already a qubit-like superposition of the two lowest Fock states in one oscillator and vacuum in the other, $|\psi\rangle_{1,0} = (0.8\ket{0}+i0.58\ket{1})\otimes\ket{0}$ shows this improvement with $R_V = 0.94$ for the gate strength amplitude $\kappa=0.46$. At the same time, more resources in the form of higher dimension of single or both oscillator states lead to significant improvement. For example, two factorized qubit-like state $U_{BS}\left(1/\sqrt{2}\right)(0.82\ket{0}+i0.57\ket{1})\otimes (0.82\ket{0}+i0.57\ket{1})$ can provide relative variance of $R_G = 0.84$ for $\kappa = 0.38$. The optimal states for other parameters can be found in Appendix B. The improvement following from the higher dimension is quantitative as well as qualitative. For higher dimensions, the relative variances $V_R$ are lower and the non-Gaussian resources surpass the Gaussian benchmark over larger ranges of the nonlinear parameter $\kappa$.
\begin{figure}
    \centering
\includegraphics[width=0.5\textwidth]{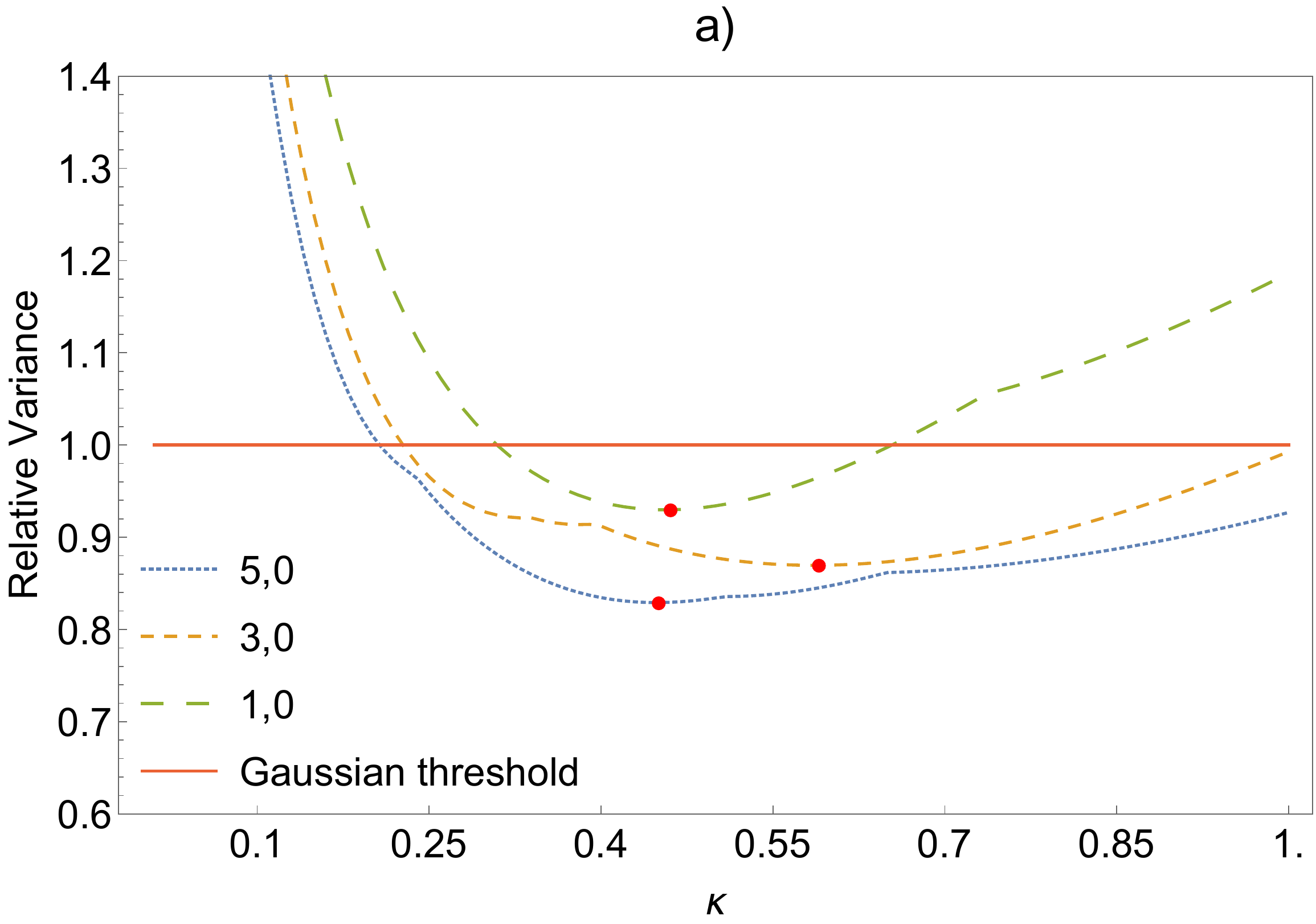}\\
\includegraphics[width=0.5\textwidth]{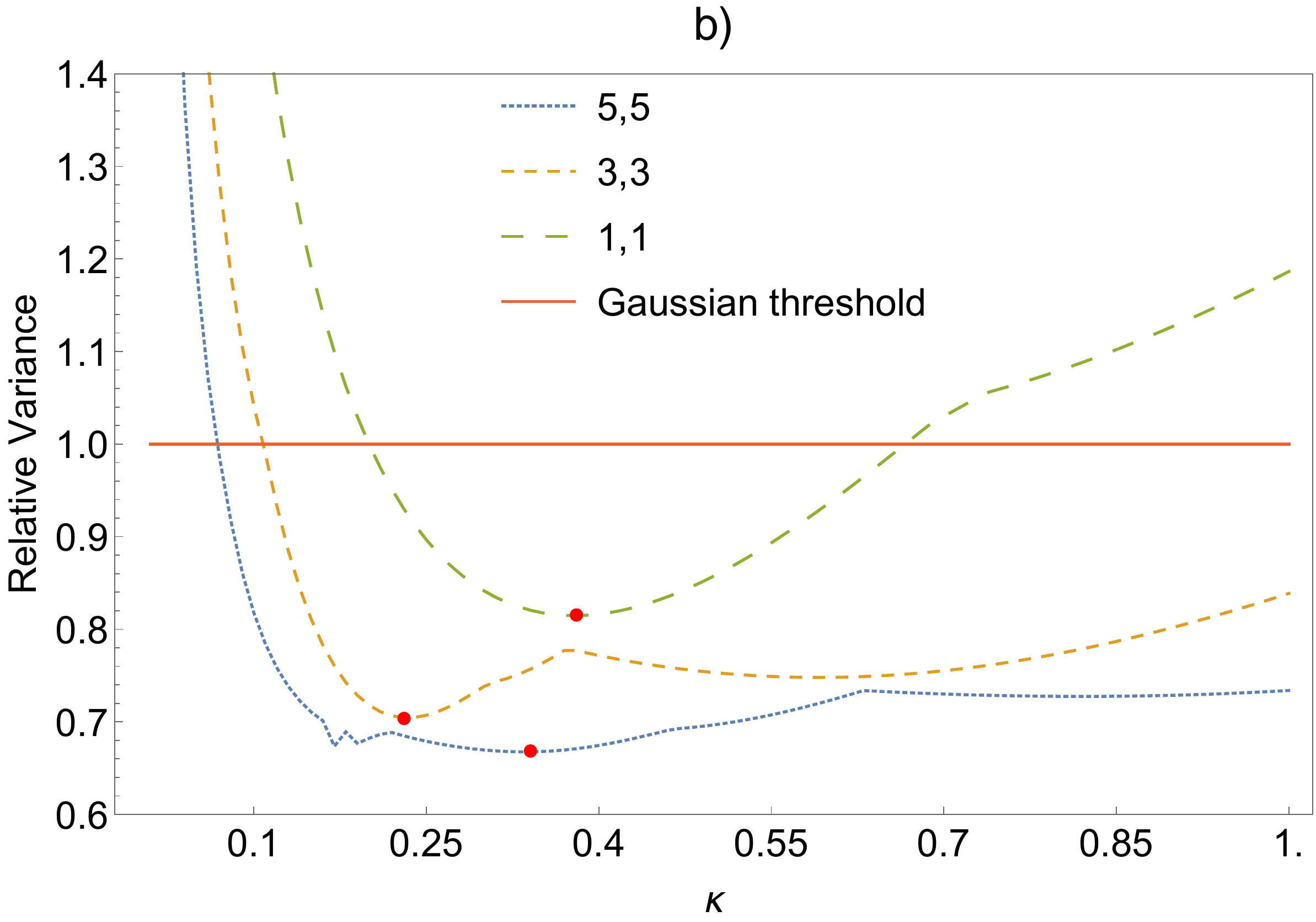}
    \caption{Comparison of relative variances $R_V$ for the simplified auxiliary state preparation setup (\ref{simplified}) with respect to different gate strengths $\kappa$. a) optimization over a state of a single oscillator with the second one being in the vacuum state b) optimization over a factorized state of the two oscillators. Minima of the relative variances are marked by red circles. The areas below the Gaussian threshold show cases where the finite non-Gaussian resource states surpass all Gaussian states.}
    \label{fig:singlemode_no_squeezing_comp}
\end{figure}

While the simplified resource states (\ref{simplified}) are interesting for proof-of-principle tests, we are ultimately interested in the best performance available. We can approach it by considering the full Gaussian processing (\ref{gaussian_processing}) as part of the state preparation (\ref{approximate_ancilla}). Interestingly, the numerical analysis reveals that the second linear coupling is not even necessary and parameter of the second beam splitter can be se to $\theta_2=0$. The squeezing operations are crucial, though. Their main role lies in adjusting the nonlinear strength $\kappa$ of the produced state. This means that a non-Gaussian core state that is optimal for one value of $\kappa$ is optimal for all of them. This gives rise to two primary invariants for each non-Gaussian core:
\begin{equation}\label{invariants}
   I_1 = \frac{\kappa}{\lambda_1^2 \lambda_2}, \quad I_2 = \frac{\kappa}{\lambda_1 \lambda_2^2},
\end{equation}
which consequently lead to secondary invariant quantities: $ \frac{\lambda_1}{\lambda_2}$, $ \frac{\kappa}{\lambda_1^3 }$, and $\frac{\kappa}{\lambda_2^3 }$. The invariants are only necessary conditions the parameters need to satisfy; not all combinations of $\lambda_1$ and $\lambda_2$ lead to an optimal output state, but
those that do obey (\ref{invariants}). This property of the Gaussian processing means that all the discussion about the resource states can be reduced to the discussion of the non-Gaussian cores.

\begin{figure}
    \centering
\includegraphics[width=0.5\textwidth]{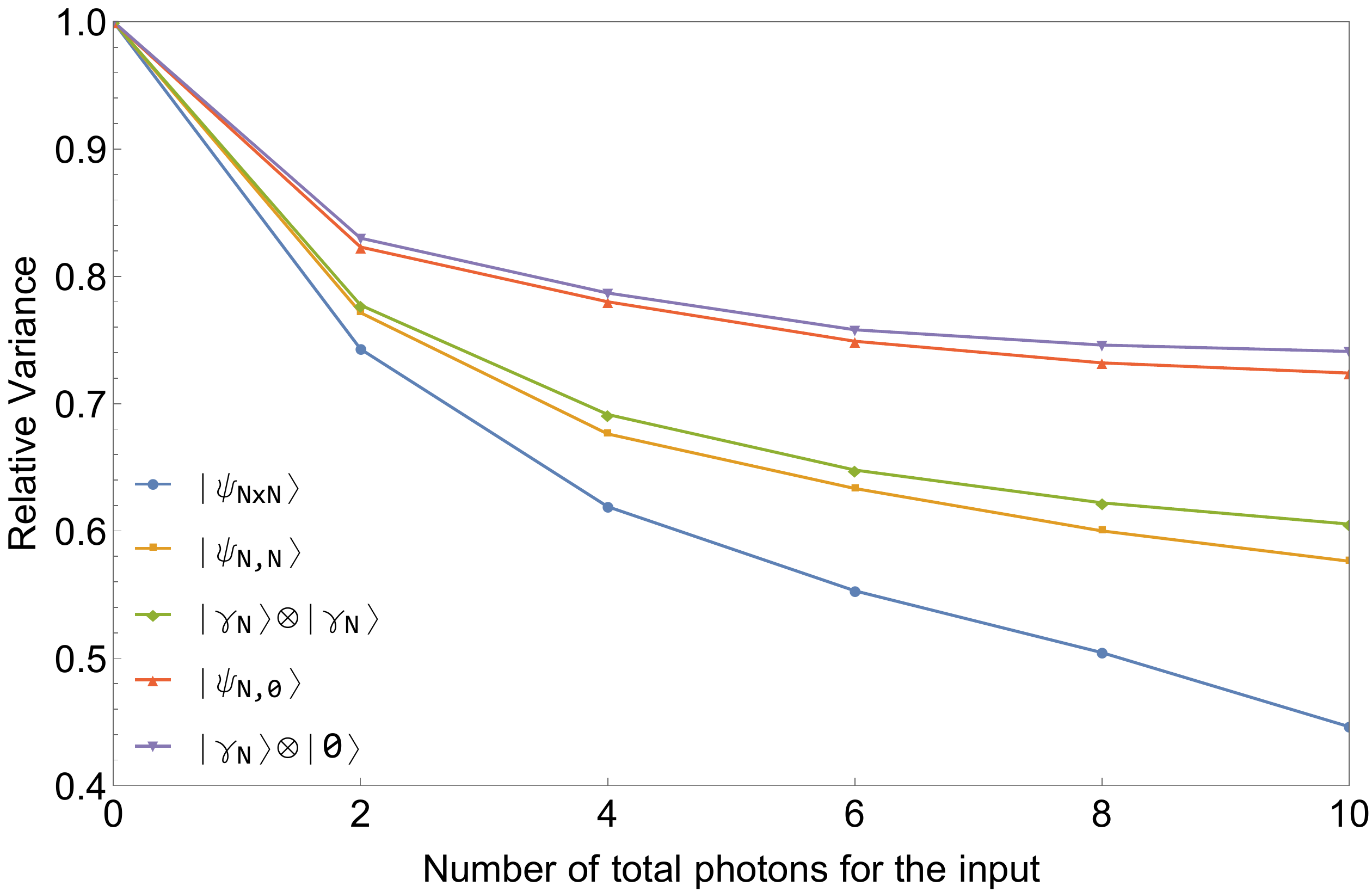}
    \caption{Relative variances $R_V$ for the optimal resource states (\ref{state_forms}), as well as for single mode cubic operation resource states $|\gamma_N\rangle$, in dependance on their total number of photons.}
    \label{fig:relative_variance}
\end{figure}

For that we compared two categories of states:
\begin{align}\label{state_forms}
|\psi_{M,N}\rangle &= \hat{U}_{G} \sum_{m=0}^M c_m |m\rangle  \otimes \sum_{n= 0}^N c'_n|n\rangle \nonumber \\
|\psi_{M\times N}\rangle &= \hat{U}_{G} \sum_{m=0}^M \sum_{n= 0}^N  c_{nm} |m\rangle |n\rangle.
\end{align}
The first one, denoted as $M,N$, consists of factorized states of the two oscillators, while the second one, marked by $M\times N$, refers to inseparable entangled states. The comparison of these classes is summarized in Fig.~\ref{fig:relative_variance}. We can see that the entangled state is more powerful than a factorized state with the same dimensions, which in turn is more powerful than the completely asymmetric case with one of the states being the vacuum state. Interestingly, even though the nonlinear coupling with Hamiltonian $\hat{H} = \kappa \hat{x}_1\hat{x}_2^2$ is not symmetric, the optimal non-Gaussian resource states $|\psi_{N,N}\rangle$ are.

The numerical methods for finding the proper non-Gaussian core states need to find the minimum of a function with a large value of variables. For example, for a state $|\psi_{N,M}\rangle$ it is $2(M+N)$ for the non-Gaussian core state alone. The optimal state $|\psi_{N\times M}\rangle$ requires even more, $2(NM + N + M)$. As we look for stronger nonlinear squeezing and the dimension of the problem increases, the numerical optimization starts becoming challenging.  One possibility is observing symmetries in the states. The observed symmetry of states $|\psi_{N,N}\rangle$, for example, reduces the number of free parameters into half. The observation that the coefficients of both separable and entangled non-Gaussian cores are, up to a $\pi/2$ phase shift, always real does that as well.
Another observation we feel worthy of mentioning is that the constituents of the found factorized non-Gaussian core states have very large overlap with the non-Gaussian resource states $|\gamma_N\rangle$ for the single mode cubic gate with Hamiltonian $\hat{H} = \kappa \hat{x}^3$ \cite{miyata2016cubic}. The overlap also translates into the nonlinear properties. This can be seen in Fig.~\ref{fig:relative_variance} where the variance of cubic resource states is very close to the variance of resource states optimized with the two-mode interaction in mind. Not only the gap vanishes when the dimension $N$ approaches infinity, but for the state $|\gamma_N\rangle\otimes|\gamma_N\rangle$ the total added variance approaches zero (see appendix \ref{sec:Limit_with_two_cubic_phase} for details). For the case of only a single cubic ancilla, $|\gamma_N\rangle\otimes|0\rangle$, the minimal achievable variance was found to be $V_R = 0.7022$, but in this case the optimality of this scenario is not guaranteed. However, the pair of perfect single mode cubic ancillas is optimal and together with Gaussian coupling it is sufficient for obtaining the perfect two-mode resource state. The nonlocal aspect of entangled core states only expedites the process.
This is very promising outcome, because the cubic phase can be produced by other platforms, like quantum optomechanics \cite{Houhou2018,Brunelli2018}, and then, they become resource for two-mode cubic couplings as well. In classical levitating optomechanics, such the cubic interaction has been already experimentally achieved \cite{Siler2017,Siler2018}.

\section{Summary and Discussion}
We have presented a general method for measurement induced implementation of multi-mode non-Gaussian interactions that couple only single quadratures of propagating modes of microwaves or light. 
The full ramifications and potential applications of these nonlinear quantum interactions are still not completely understood, mainly because the difficulty accompanying both the theoretical analysis the potential experimental implementation. We believe that our method can be used for first proof-of-principle experimental tests which will in turn stimulate the theoretical investigation of this sizeable yet unexplored area.

The proposed method is based on multi-mode extension of the original GKP scheme \cite{GPK}. It relies on a non-demolition non-Gaussian measurement \cite{miyata2016cubic} of each of the individual modes followed by a suitable joint nonlinear feed-forward operation. The method can be also translated to a loop architecture of nonlinear gates on temporal propagating modes \cite{Takeda2017}.
The crucial element of the implementation is the resource state. For $N$-mode operations, the resource state is an entangled state simultaneously squeezed in $N$ nonlinear nonlocal quadrature operators. The ideal state is a joint eigenstate of these operators but, similarly to single quadrature eigenstates, it can be approached only asymptotically.
We have investigated techniques for finding the required state for the example of two-mode cubic gate with Hamiltonian $\hat{H} = \kappa \hat{x}_1 \hat{x}_2^2$. First of all, we have found that for some values of the nonlinear parameter $\kappa$ the entangled resource outperforming purely Gaussian approximation of the coupling can be obtained simply by creating a suitable superposition of Fock states $|0\rangle$ and $|1\rangle$ and splitting them on a beam splitter. The squeezing of the nonlocal quadratures can be then significantly improved either by using higher Fock states, or by interfering a pair of such states on the beam splitter.
Moreover, singe-mode cubic phase states can be merged in the Gaussian two-mode couplings to prepare required entangled resource states for two mode-cubic phase gate. This opens possibility for other platforms, like nonlinear quantum optomechanics, to deliver the required resources for nonlinear optical circuits.
All these techniques are feasible with the available experimental techniques and the nonlinearity and its properties can therefore be readily experimentally tested.

Universal ancillary states can be obtained by preparing a specific single or multi-mode finite superposition of Fock states and altering it by suitable Gaussian coupling which includes squeezing operations. The squeezing operations can be used for tuning the value of parameter $\kappa$ so the underlying Fock superposition state, the non-Gaussian core of the resource state, is universal for all $\kappa$ and depends only on the selected dimension. Its exact form needs to be found by numerical methods and the nonlinear squeezing significantly depends on the complexity: entangled superposition of Fock states leads to a better resource state than factorized one. This is an intriguing example of genuine non-Gaussian entangled state significantly outperforming state which is also non-Gaussian but with entanglement created only by Gaussian operations. This opens up an interesting avenue of research: while the conditional techniques for preparation of single mode superpositions of Fock states are well known \cite{Dakna1999, Yukawa2013,Pfaff2017}, preparation of multi-mode state has been devised only for specific classes of states \cite{Shahandeh2016} and the general case remains and open problem even in the conditional regime. The deterministic preparation of such states then arises as the ultimate challenging goal of future investigations. 


\section*{Acknowledgements}
We were supported by Project No. GB14-36681G of the Czech Science Foundation.
P.M. acknowledges LTC17086 of INTER-EXCELLENCE program of Czech Ministry of Education. R.F. have received national funding from the MEYS of the Czech
Republic (Project No. 8C18003) under Grant Agreement No. 731473 within the QUANTERA ERA-NET cofund in
quantum technologies implemented within the European
Unions Horizon 2020 Programme (Project TheBlinQC).

\appendix
\section{Minimization method}\label{sec:Minimization_method}

   The value that we are trying to minimize is:

     \begin{equation}\label{eq:var_for_O}
 \bra{\psi}  A (a,a^\dag)\ket{\psi}-\left(\bra{\psi}B  (a,a^\dag)\ket{\psi}\right)^2 \quad s.t. \braket{\psi|\psi}=1
 \end{equation}

 Here $A$ and $B$ are known operators which are represented as matrices in this particular vector space that we are limited to. The vector that minimizes this value can be expanded in an orthogonal basis,

 \begin{equation}
 \ket{\psi}=c_0\ket{0}+c_1\ket{1}+...c_f\ket{f}
 \end{equation}

 Plugging it into eq. \eqref{eq:var_for_O} one can derive the following quartic polynomial that has to be minimized over a complex unit sphere:

 \begin{equation}
 \sum_{nm}A_{nm}c_n^*c_m-\left(\sum_{nm}B_{nm}c_n^*c_m\right)^2
 \end{equation}

 For calculating the variances following matrix elements has to be calculated:

 \begin{equation}
 \bra{\psi} A (a,a^\dag)\ket{\psi}, \quad \bra{\psi} B (a,a^\dag)\ket{\psi}
 \end{equation}

 We have calculated the matrix elements in the following way:

 \begin{equation}
 A_{mn}=\bra{m} A (a,a^\dag)\ket{n}, \quad  B_{mn}=\bra{m} B (a,a^\dag)\ket{n}
 \end{equation}

Here $m$ and $n$ are the Fock states, i.e. $a^\dag a \ket{n}=n\ket{n}$ However note that since we are dealing with complex numbers the number of parameters is two times the dimension of the vector space. (Due to normalization condition it is one less than two times the dimension to be more precise.) The most convenient way to deal with such system is to expand the coefficients into real and imaginary parts and minimize the equation in real vector space.

Eventually this method relies on minimization of a quartic polynomial over a unit sphere that contains large number of variables which is computationally demanding. This is actually a NP-hard problem \cite{luo2010semidefinite}. We have used a particular built in minimization algorithm from Mathematica software. More specifically we have employed a quasi newton type BFGS algorithm through using tens of thousands of initial points for the search.

\section{Table of the optimal non-Gaussian states and respective Gaussian parameters}
In this appendix we present some optimal input states and the corresponding Gaussian transformation. Note how the two mode states are symmetrical under the exchange of inputs. Also the coefficients are all real up to a $e^{\frac{i \pi}{2}(n_1+n_2)}$ operation.
For setups that contain only passive elements:
\begin{itemize}
    \item When limited to single photon in the first input for $\kappa=0.46$
    \begin{equation} U_{BS}(0.86)(0.8\ket{0}+i0.58\ket{1})\otimes \ket{0}
\end{equation}
\item When limited to two photons in the first input For $\kappa=0.38$
\begin{equation}
  U_{BS}(0.87)(0.47\ket{0}+i0.78\ket{1}-0.4\ket{2})\otimes \ket{0}
\end{equation}
\item When limited to one photons in both inputs for $\kappa=0.38$
    \begin{equation}
    U_{BS}\left(\frac{\pi}{4}\right)(0.82\ket{0}+i0.57\ket{1})\otimes (0.82\ket{0}+i0.57\ket{1})
\end{equation}
\item When limited to two photons in both inputs for $\kappa=0.29$
    \begin{equation}
    U_{BS}\left(\frac{\pi}{4}\right)(0.51\ket{0}+i0.76\ket{1}-0.38\ket{2})\otimes (0.51\ket{0}+i0.76\ket{1}-0.38\ket{2})
\end{equation}
\end{itemize}

For setups that employ the full Gaussian processing with passive coupling and single mode squeezing operations:

\begin{itemize}
    \item When limited to single photon in the first input
\begin{equation}
    U_{BS}(-0.16)S_1(1.04)S_2(1.47)U_{BS}(1.07)(0.8\ket{0}+i0.59\ket{1})\otimes\ket{0}
\end{equation}
\item When limited to two photons in the first input \begin{equation}
U_{BS}(-0.22)S_1(1.04)S_2(1.47)U_{BS}(1.14)(0.47\ket{0}+i0.77\ket{1}-0.41\ket{2})\otimes\ket{0}
\end{equation}
\item When limited to one photons in both inputs
    \begin{equation}
    S_1(1.06)S_2(1.58)U_{BS}\left(\frac{\pi}{4}\right)(0.7\ket{00}+i0.42\ket{01}+i0.42\ket{10}-0.38\ket{11})
\end{equation}
\item When limited to two photons in both inputs.
    \begin{equation}
    \begin{split}
    S_1(1.07)S_2(1.79)U_{BS}\left(\frac{\pi}{4}\right)&(0.33\ket{00}+i0.36\ket{01}-0.11\ket{02}\\
    &+i0.36\ket{10}-0.59\ket{11}-i0.3\ket{12}-0.11\ket{20}-i0.3\ket{21}+0.21\ket{22})
    \end{split}
\end{equation}
\end{itemize}

\section{Achieving minimal variance using cubic phase states.}\label{sec:Limit_with_two_cubic_phase}

In this section we show that using prefect cubic phase states with Gaussian resources one can approach two mode cubic gate in principle.

We first generate two single mode cubic phase gate ancillae and then couple it using a fifty-fifty beam splitter:
\begin{align}
U_{BS}\int e^{itX_1^3} \ket{x_1}dx_1\int e^{itX_2^3} \ket{x_2}dx_2 \nonumber \\
= \int e^{it'(X_1-X_2)^3}e^{it'(X_1+X_2)^3} \ket{x_1}\ket{x_2}dx_1dx_2
\end{align}
After that we can implement squeezing operations in modes 1 and 2:
\begin{equation}
\int e^{it'(2X_1^3\lambda_1^3+6X_1X_2^2\lambda_1\lambda_2^2)} \ket{x_1}\ket{x_2}dx_1dx_2
\end{equation}
If we choose the parameters so that $\lambda_1\lambda_2^2=1$ and $\lambda_1 \rightarrow 0$, only the two-mode term remains relevant and the ideal resource state is obtained.


\end{document}